# A mechanistic model of trust based on neural information processing

**Scott E. Allen**
ORCID ID: 0000-0002-7399-3408
Department of Physics, Cornell University, Ithaca NY 14853
Current address: Department of Physics, MIT, Cambridge, MA 02139
scallen@mit.edu

**René F. Kizilcec**
ORCID ID: 0000-0001-6283-5546
Department of Information Science, Cornell University. Ithaca NY 14853
kizilcec@cornell.edu

**A. David Redish** *Corresponding author*
ORCID ID: 0000-0003-3644-9072
Department of Neuroscience, University of Minnesota, Minneapolis MN 55455
redish@umn.edu

## Abstract

Trust is central to human social interactions, manifesting as a critical information processing step in taking actions that make one vulnerable to another. We argue that trust depends on the decision-making processes that arise in neural systems. Building on advances in the cognitive neuroscience of decision making, we propose a mechanistic model of trust arising differently in multiple parallel systems that perform distinct, complementary information processing. Because each system learns via different computational mechanisms, they will interact with the environment differently, and trust can be created (or destroyed) in multiple ways. This systems-level taxonomy of information representations provides a principled basis for differentiating forms of trust, linking them to specific learning processes, and generating testable predictions about their expression in behavior. Furthermore, because these different computational processes are implemented by different neural circuits, our theory makes testable predictions about the different neural circuits underlying different kinds of trust. By situating trust within a broader theory of neural decision systems, our account unifies diverse findings across psychology, neuroscience, and the social sciences, and offers a foundation for explaining how humans develop, maintain, lose, and repair trust in a complex social world.

## Introduction

Societies are built on trusting relationships, and institutions function only when their members decide to rely on each other (Cook et al., 2013; Dawes, 1980; B. Hare & Woods, 2021; Luhmann, 2018; Ostrom, 1990; Redish, 2022; Simpson, 2007; Stout, 2010). The question of the definition of the term "trust" is a long and contentious one, but most researchers agree that trust is an important component in decisions to take actions that make one vulnerable to others (Bigley & Pearce, 1998; Cook et al., 2007; Davis & Whinston, 1962; F. Krueger & Meyer-Lindenberg, 2019; Luhmann, 2018; Rousseau et al., 1998). Many actions potentially leave one vulnerable to another's choices, whether that be physically, financially, emotionally, or socially. However, if one has prior information that the risk of betrayal is low (because one has observed that person's actions in the past, because there are other signals suggesting betrayal is less likely to occur, or because there are mitigating factors that would prevent betrayal or mediate its consequences), then one can become more willing to take that potentially vulnerable action.

We argue that trust is multifarious, and that each form of trust is best understood as arising from separable decision-making systems, shaped by the neural systems that govern how people learn, store, and process information. Contemporary neuroscience shows that multiple, interacting decision-making systems guide behavior, each with distinct representational formats and learning mechanisms (Redish, 2013). Building on this foundation, we propose a mechanistic model of trust that links these neural systems to different forms of trust, offering a new taxonomy that generates specific predictions about how trust is formed, maintained, lost, and repaired.

In this manuscript, we start from the definition of trust as the step in the information process underlying decision-making that makes one more likely to take a cooperative action with potential for vulnerability. We note that many researchers confine the term "trust" to specific aspects of specific decision processes (such as only when it arises from an assessment of the conspecific's character (Cook et al., 2007; Mayer et al., 1995), or only when it arises from calculated decisions (Gambetta, 1988; Gambetta & Hamill, 2005), see Bigley and Pearce (1998) for an in-depth review of this debate. In contrast, but consistent with Bigley and Pearce (1998) and other syntheses (Rousseau et al., 1998), we will not make this restriction. Instead, we will define trust functionally (as this step in the information process) and follow that functionality to outline the implications.

Importantly, we do not assume that the assessment of risk is correct — many people "trust" the untrustworthy and take actions that do lead to betrayal. Also, importantly, we do not assume that this assessment is explicit, cognitive, or conscious. We do not assume that the assessment is an actual calculation of the risk. Nevertheless, this definition of trust shifts the question from *What is trust?* to one of *Why does the agent take the action it does?* This definition of trust moves the concept of trust from its own philosophical intension to one of decision-making.

Decisions emerge from integrating information about the present (perceptions), the past (memory), and the future (goals, motivations, and expectations). In the case of trust, expectations are especially critical, since trust requires anticipating the future actions of others within one's social sphere. The way information about that social sphere is represented, how that information is transformed as the brain computes potential actions, and how it is updated through learning from observed outcomes of actions all shape behaviors such as trust, cooperation, and adaptation. Modern theories of decision making emphasize that these processes are carried out by distinct but interacting neural systems, each with its own modes of information representation, computation, and learning. Specifically, each system has identifiable modes of representing, computing, and learning expectations of others' future actions. This systems-level taxonomy of decision making therefore provides a direct foundation for developing a taxonomy of trust, and yields implications for how trust is developed and enacted.

## Multiple decision-making systems imply a taxonomy of trust

Current neuroscience suggests that decision-making relies on multiple neural circuits, each processing information differently and enabling different psychological resources. Foundationally, these theories arise from the computational neuroscience assumptions that neural circuits are implementing computational processes (P. S. Churchland & Sejnowski, 1994; Dayan & Abbott, 2001; Marr, 1991), and that decision-making is a computational process (Glimcher et al., 2008; Kahneman & Tversky, 2000; Lieder & Griffiths, 2019; Rangel et al., 2008; Redish, 2013; D. V. Smith et al., 2026). These two hypotheses suggest that different neural circuits implement separable decision processes performing different computations. One of the most important discoveries in the computational sciences is that how information is represented determines what can be done with it, and thus different representations enable different transformations, which change both the speed and functionality of that information processing (Cormen et al., 1990; T. M. Mitchell et al., 1986) — this has obvious consequences for decision-making (Glimcher et al., 2008; Lieder & Griffiths, 2019; K. J. Mitchell, 2023; Redish, 2013; D. V. Smith et al., 2026). In particular, how each system represents information includes how it represents any expectations of the future, which limits each system to certain computations during decisions, including those involving social risk.

Modern neuroscientific theories of decision-making typically identify three key action-selection systems, each with distinct fundamental properties and each driven by different neural circuits. Although they often go by different names across different psychological and neuroscientific literatures,[1] these three key systems can be identified by their computational processes — Deliberative planning systems based on serial sampling, Instinctual association systems based on releasing actions from a limited repertoire in learned situations, and Procedural habit systems based on cached associations of arbitrary learned action sequences taken in learned situations (Gesiarz & Crockett, 2015; Rangel et al., 2008; Redish, 2013; Redish et al., 2008; van der Meer et al., 2012).

[1] Deliberative decision-making is also known as model-based (Doll et al., 2015; S. W. Lee et al., 2014; Niv et al., 2006), action-outcome (Dickinson, 1980; Domjan, 1998), goal-directed (Dickinson & Balleine, 1994; Killcross & Coutureau, 2003a), locale navigation (O'Keefe & Nadel, 1978; Redish, 1999), and serial sampling (Stewart et al., 2006). Instinctual decision-making is also known as Pavlovian (Domjan, 1998; Gesiarz & Crockett, 2015; Rescorla, 1988), stimulus-stimulus (Dickinson, 1980), state-based (Dayan & Balleine, 2002; LeDoux & Daw, 2018), and incentive learning (Balleine, 2004; Berridge & Robinson, 2003). Procedural decision-making is also known as model-free (S. W. Lee et al., 2014; Niv et al., 2006), stimulus-action (Domjan, 1998; Hull, 1943; Mackintosh, 1974), route navigation (O'Keefe & Nadel, 1978; Redish, 1999), and habit-learning (Dickinson, 1985; Graybiel, 2008; Killcross & Coutureau, 2003b; K. S. Smith & Graybiel, 2013).

Each system has unique functionalities, limitations, and learning mechanisms. Thus we propose that trust manifests differently within each of these systems.

- The *Deliberative System* uses structured maps of cause-and-effect to generate simulations of future events (Niv et al., 2006; Redish, 2016; Samborska et al., 2022). Trust based on deliberative resources would use explicitly imagined predictions about the consequences of the interactions between one's own actions and those of other agents.
- The *Instinctual System* uses learned associations to release actions from a limited repertoire of possibilities related to the ethology of the individual's species (Breland & Breland, 1961; Dayan et al., 2006; LeDoux & Daw, 2018). In humans, a key subset of this repertoire involves social interactions, particularly those that establish community (Brewer & Kramer, 1986; Crockett, 2017; Marsh, 2017; Redish, 2022). Trust based on instinctual resources will generate social connections such as tribal similarities and affective emotions.
- The *Procedural System* relies on learned pattern recognition processes to recognize situations and then to release well-practiced action sequences aligned with key dimensions of those situations producing automated habits (Ericsson et al., 2018; Graybiel, 2008; Greenstreet et al., 2025; Jog et al., 1999; Klein, 1999). It allows for precise, rapid performance of an arbitrarily learned action in an arbitrarily learned situation, but requires repetitive practice to learn. Trust based on procedural resources will generate precise and repeatable cooperative behaviors.

Each of these decision-making resources depend on representations of different kinds of information transformed in different ways, and therefore function as different kinds of expectations of future events. Since each system holds expectations independently of the others, trust can be decomposed into *deliberative trust*, *instinctual trust*, and *procedural trust*. We argue that these three distinct forms of trust work in tandem to yield the cooperative behaviors upon which our human relationships and institutions are built. Although these are not the only decision computational mechanisms that have been proposed, they are the primary ones that are consistently identified in nearly all neurophysiological decision models, and thus can serve as a starting point for a discussion of the neurophysiology of trust.

## Deliberative trust

*Deliberative decision-making* involves explicitly running a mental simulation of one's own actions, the actions of others, and the world itself, to identify the outcomes that will likely follow. The key benefit of deliberative decision making is the ability to make novel choices in new situations, provided one can predict the consequences of those actions (Niv et al., 2006; Redish, 2013, 2016). Computationally, deliberative decision systems entail a search-by-sampling process (Hunt et al., 2021; Redish, 2016; Stewart et al., 2006; van der Meer et al., 2012) in which a potential future outcome is explicitly imagined and then evaluated. Fascinatingly, these search processes seem to follow depth-first rather than breadth-first algorithms, in which sequences of scenarios are created and then evaluated (Faulkner et al., 2021; Hunt et al., 2021; Huys et al., 2012). See Figure 1.

Figure 1: Deliberative decision making.

KEY FACTORS:

- Information is represented in high-dimensional maps, including cause-and-effect relationships.
- Expectations are explicit mental simulations of future events.
- Strategically plans out cooperative actions based on imagined outcomes.

COMPUTATION: The deliberation system works through a depth-first search-by-sampling process. For example, in making the decision of whether to take action A1 or A2 in state S0, the deliberation system would run sequences through the consequences (here A1 leads to S1 to S2 to S3 which can then be evaluated using V(S3), while A2 leads to S4 and then through S5, S6, and S7 which can then be evaluated. Depending on the evaluation of these expected subsequent states, the decision will be made whether to take A1 or A2. Note that the deliberation system runs through the complete path S1 to S2 to S3 before considering S4, which makes this a depth-first rather than a breadth-first search process.

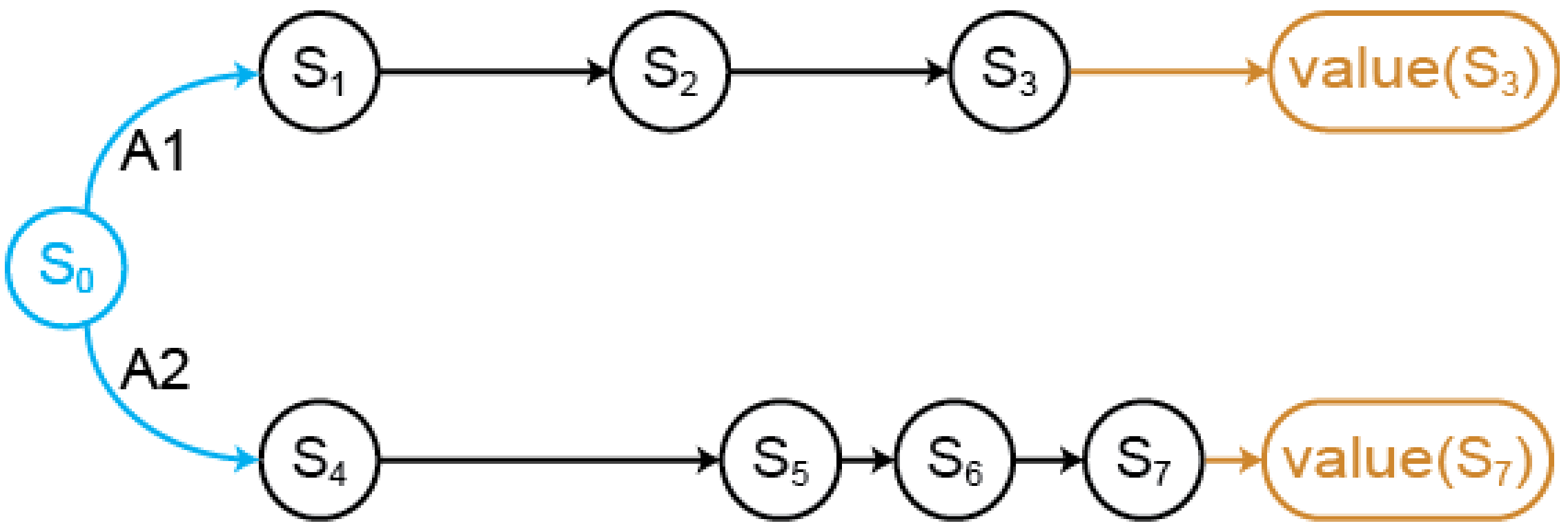

Neurophysiologically, the deliberative system is based on a complex neural circuit involving the prefrontal cortex, the hippocampus, the ventral aspects of striatum (particularly nucleus accumbens core), and the orbitofrontal cortex. The prefrontal cortex provides goal-setting functionality that seems to provide strategic representations. In humans and other primates, deliberation depends on dorsolateral cortex (Fellows & Farah, 2005; Huang et al., 2014; Passingham, 1993; Tanaka et al., 2008; Wise, 2008), but in rodents (who do not have a dorsolateral cortex, (Laubach et al., 2018; Wise, 2008)), the more medial aspects of prefrontal cortex seem to play these roles (Diehl & Redish, 2023; Powell & Redish, 2016; Sharpe et al., 2015; Tang et al., 2021). In humans, deliberative decision-making is associated with activation in the prefrontal executive and default mode networks (Buckner & Carroll, 2007; Schacter et al., 2007). The prefrontal network is particularly active in goal transitions (Schuck et al., 2015), and lesions to that prefrontal network impair transitions between goals and strategy (Milner, 1982).

The hippocampus plays a role in episodic future thinking across species (Clayton et al., 2008; Comrie et al., 2022; Hassabis & Maguire, 2011; Johnson & Redish, 2007; Schacter et al., 2007).

In humans, in addition to the well-established inability to remember episodic past events after hippocampal lesions (Cohen & Eichenbaum, 1993; Nadel & Moscovitch, 1997), it has also been shown that human patients with hippocampal lesions are unable to imagine potential episodic futures (Hassabis et al., 2007). In rodents, hippocampal representations sweep forward down potential paths during deliberative choices (Johnson & Redish, 2007; Tang et al., 2026; Yu et al., 2026; Zheng et al., 2021; Zutshi et al., 2025), suggesting a neurophysiological role in planning (Damphousse, Gagliardi, et al., 2026; Redish, 2016) — these representations are dependent on proper functioning of the prefrontal cortex (Schmidt et al., 2019) and couple to prefrontal cortex during deliberation (Benchenane et al., 2010; Hasz & Redish, 2020; Hyman et al., 2010; Jones & Wilson, 2005; Mugan, Hoffman, et al., 2024; Tang et al., 2021, 2026).

For deliberative decision making, it is not enough to imagine future outcomes, one also needs to evaluate them. There is extensive evidence in both rodents and primates (including humans) that the nucleus accumbens core (which forms part of the ventral aspects of the striatum) and the orbitofrontal cortex are critical components of that evaluation step (T. A. Hare et al., 2008; Knudsen & Wallis, 2022; McDannald et al., 2011; Padoa-Schioppa & Assad, 2006; Sharpe et al., 2015; van der Meer et al., 2010). In both rodents and non-human primates, neural activity in these areas encodes value-related signals (Cunningham & Redish, 2025; Padoa-Schioppa & Assad, 2006; Roesch et al., 2009; Stott & Redish, 2014; Zhou et al., 2019) tied to the hippocampal representations (Balewski et al., 2023; Riceberg et al., 2022; van der Meer & Redish, 2011). Fascinatingly, these value signals transiently reflect individual options consistent with a sampling hypothesis (Balewski et al., 2023; Rich & Wallis, 2016; A. Steiner & Redish, 2012; Stott & Redish, 2014; van der Meer & Redish, 2009). In humans, damage to the orbitofrontal cortex in particular leads to problems with value-based decisions, particularly those that involve direct comparisons between complex options (Fellows & Farah, 2005; Henri-Bhargava et al., 2012).

In general these neural circuits enable the imagination and evaluation of alternative outcomes, a process referred to as “mental time travel” (Buckner & Carroll, 2007; Clayton et al., 2008; Johnson et al., 2009; Redish, 2025; Suddendorf, 2013; Suddendorf & Corballis, 2007; Trope & Liberman, 2010). Similar circuits enable *episodic future thinking*, which uses causal expectations of the consequences of behavior to predict future outcomes (Hassabis & Maguire, 2011; Redish, 2016; Schacter et al., 2007), *counterfactual thinking*, which uses causal expectations to infer what would have happened if an alternative choice had been made, which is critical to the phenomenon of regret (Boorman et al., 2009; Coricelli, 2025; Kocharian et al., 2025), as well as *episodic memory*, which entails a recreation of past events (Cohen & Eichenbaum, 1993; Crystal, 2024; Nadel & Moscovitch, 1997; Schacter, 2001; Squire, 1987; Suzuki & Clayton, 2000), and may play roles in both rumination and PTSD (Perl et al., 2023; Puccetti et al., 2025).

By storing information about causal relationships, the Deliberative System supports cooperative choices in situations where those relationships are understood. This system is unique in that expectations (information about the future) exist explicitly as mentally simulated events. This enables people to take cooperative actions they have never taken before and accept vulnerabilities that are unfamiliar, such as navigating a complex legal interaction or confronting a coworker about an integrity violation. For example, being told that violating a trust will lead to legal consequences can allow one to express deliberative trust and make oneself vulnerable to another — because one knows the other will face legal consequences if they violate that vulnerability. The Deliberative system learns by extracting causal information patterns from the environment, including the social environment. We propose that learning trust as a deliberative resource in this same manner enables flexible adaptive cooperation.

Deliberative trust has the same limitations as all other phenomena that depend on deliberative resources. First, deliberation is cognitively tiring. Second, deliberation is inherently a depth-first search process, even when a breadth-first approach would be more effective. Deliberation includes both deep planning that traces a chain of actions to its predicted conclusion, and also shallow planning that only looks a few steps ahead or only considers some possibilities. Though taking time to carefully deliberate can mitigate some of the tendency to jump to conclusions or to choose a seemingly obvious option, deep deliberation can also lead someone to consider rare possibilities that become emotionally important. Deep deliberation shares many of the limitations of shallow, heuristic deliberation, including incomplete awareness of possibilities and limited evaluative capacity. Since deliberation is a depth-first search, deep deliberation is also sensitive to initial impulses just like shallow deliberation. Likewise, shallow deliberation can still generate novel actions in unfamiliar situations, selecting them on newly evaluated outcomes. Whether navigating complex simulations of the future or relying on simple heuristics, deliberative trust arises from the same fundamental causal information representation mechanism and must be learned through attention to causal relationships.

## Instinctual Trust

*Instinctual trust* involves unlearned cooperative actions. All mammals have a genetic repertoire of innate unlearned somatic and affective actions that they learn when to release. In humans, a large part of the instinctual repertoire promotes social bonding and group belonging through learning signals related to community (Gesiarz & Crockett, 2015; Marsh, 2017). Computationally, instinctual decision systems entail learned associations between cues and conditions and somatic and emotional responses (Breland & Breland, 1961; LeDoux & Daw, 2018; Lindquist et al., 2022). These associations reflect the information that the cues provide to differentiate one condition from another (Balleine, 2004; Domjan, 1998; Gallistel, 1990). From these associations, the animal can release appropriate behaviors from a limited repertoire (Breland & Breland, 1961; LeDoux & Daw, 2018; Rescorla, 1988). See Figure 2.

Neurophysiologically, the instinctual system is primarily associated with the amygdala (Janak & Tye, 2015; LeDoux & Daw, 2018; Marsh, 2017; Seymour & Dolan, 2008), where the connection strength between (baso)lateral amygdala to central amygdala has been directly identified with the learned associative strength between the cues and actions taken (Nabavi et al., 2014; Rogan et al., 1997). Simple responses are driven by amygdala projections to the periaqueductal gray, a small area in the basal midbrain, while more complex responses are driven by amygdala projections to the nucleus accumbens shell (which forms a part of the ventral striatum, (Bandler & Shipley, 1994; Lind et al., 2023; Mobbs et al., 2007; Ramirez et al., 2015)). In primates in particular, amygdala neurons are specifically sensitive to social cues, differentiating (for example) touch from objects and social companions (Gothard & Fuglevand, 2022). In humans, amygdala activity is strongly coupled to group membership, differentiating between in-group and out-group interactions (Edmonds et al., 2024; Marsh, 2017) and is deeply involved in recognizing trust interactions in experiential tasks (Engell et al., 2007; Koscik & Tranel, 2011; Stanley et al., 2012; Todorov & Engell, 2008). These socially-learned signals provide for fast, implicit recognition of signals, often involving simple factors such as facial recognition (Engell et al., 2007; Stanley et al., 2012; Todorov & Engell, 2008). Importantly, these signals provide for very fast definitions of ingroup and outgroup and can produce fast negative (and potentially problematic) responses to outgroup identifications that can only be overridden by other systems or through extended learning (Hughes et al., 2017; Redish, 2022; Stanley et al., 2012; Van Bavel et al., 2008).

Figure 2: Instinctual decision making.

KEY FACTORS:

- Information is represented as states of the world in which a limited repertoire of actions can be released.
- Expectations are implicit within group membership.
- Generates affective states of social belonging. Trust and distrust arises out of parochial altruism — cooperation within the group and non-cooperation outside of it.

COMPUTATION: The instinctual system associates cues (S0) with conditions in the world and somatic states in which certain actions are reasonable to take (S1). These actions come from a limited repertoire appropriate to the condition. (Top: before learning. Bottom: after learning.) The associated relationship reflects how much information S0 provides that S1 is also true and enables S0 to activate S1 and release the appropriate actions.

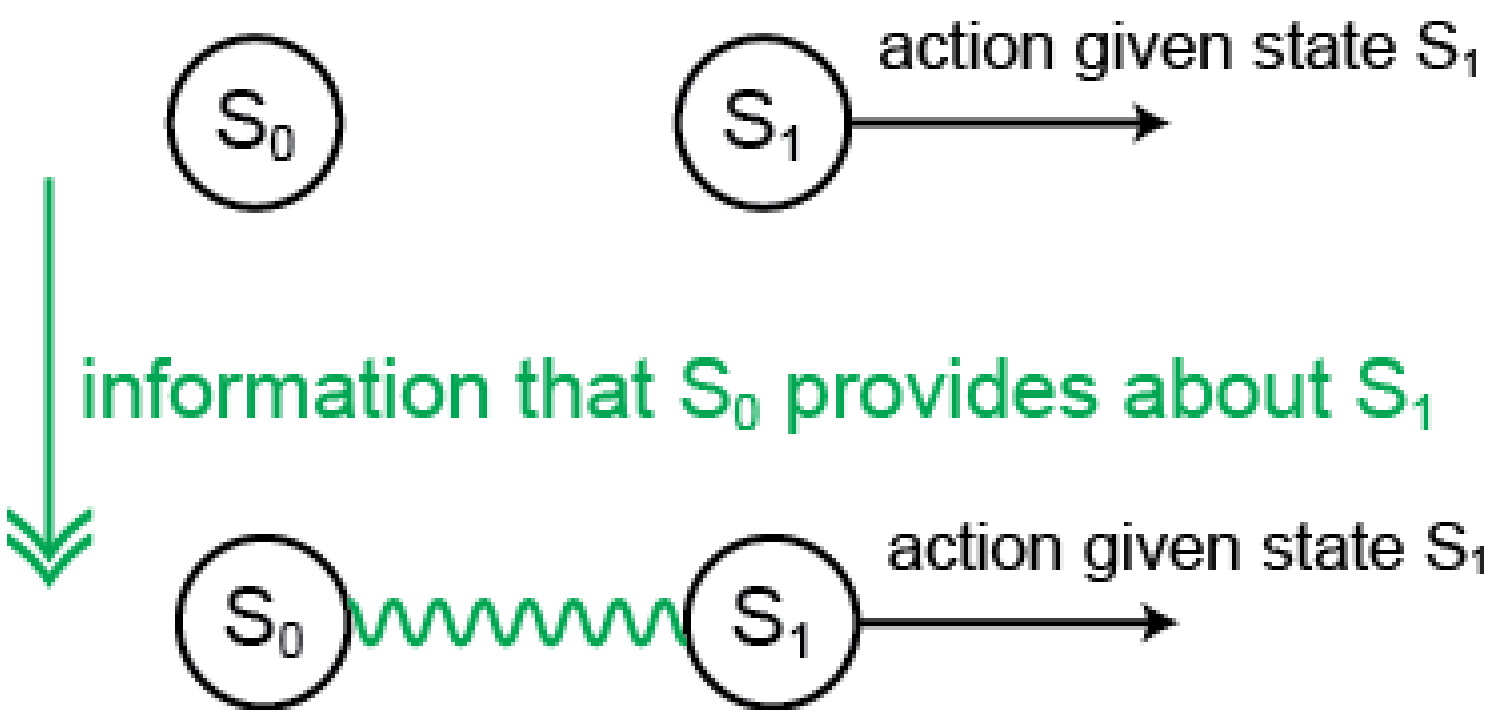

Cortically, the key structures for instinctual responses tend to be driven by the salience network, consisting primarily of anterior insula and subgenual anterior cingulate cortex. Anterior insula is often found to be active after emotional responses to social betrayal, such as offers that are too low in the ultimatum game (Sanfey et al., 2003) or inadequate investment returns in the investment or trust game (King-Casas et al., 2005, 2008). Fascinatingly, empathy entails activation of this network, including anterior insula, the subgenu anterior cingulate cortex, and amygdala (Bellucci et al., 2020; Lockwood, 2016; Singer, 2008), in contrast to simple prediction of another's behavior, which activates more deliberative circuits such as dorsolateral prefrontal cortex and social recognition structures such as the temporal parietal junction (TPJ) (Bellucci et al., 2020). In particular, thinking about a friend's pain produces activation in salience network circuits (anterior insula, subgenual anterior cingulate cortex) that thinking about a stranger's pain does not (Eisenberger, 2012).

In primates in general and humans in particular, the Instinctual System attends to signals which predict group membership — group membership predicts cooperation (Brewer, 2008; Brewer & Kramer, 1986; B. Hare & Woods, 2021; Redish, 2022). Importantly, the data suggests that the Instinctual System does not store any explicit information about these cooperative relationships, such as constructs of benevolence or reciprocity. Rather, it accesses a set of predictors that have generally been evolutionarily predictive of cooperative relationships. Individually these predictors may be non-structural, but together they enable complex social dynamics to emerge through social learning. Thus, instinctual resources are learned through repeated exposure to predictors of in-group cooperation. While it can take time to develop a new group identity, one can signal membership in an existing group rapidly. This dependence on group identity and membership leads to unique functions and limitations of *instinctual trust.*

The Instinctual System relies on group membership as a powerful heuristic to predict safety and cooperation. It underlies the parent-child relationship and the general importance of kith (community) and kin (family) (Baumgartner et al., 2008; Marsh, 2017; Tomasello, 2016). *Instinctual trust* is, in part, the expectation of group membership. Analyses of the non-zero-sum economics of cooperation suggest that an important component leading to cooperation is the expectation of repeated interactions, whether that be individual reciprocal interactions or through reliable third-party chains (Axelrod & Hamilton, 1981; Redish, 2022; Skyrms, 2004; Sober & Wilson, 1998; D. S. Wilson, 2015). Group membership can serve as a proxy for the expectation of these repeated interactions (Fehr & Fischbacher, 2003; Redish, 2022; Sober & Wilson, 1998; D. S. Wilson, 2015). If a group rejects treacherous members (Boehm, 2012), its members can generally be treated as safe to trust, making the extensive calculations required for deliberative trust unnecessary.

These instinctual processes signaling group membership are not without limitations. Group membership does not necessarily imply that one should trust the others within the group. There are many examples where group membership can lead to errors in judgment, such as an unwillingness to reject flawed decisions made by the group (Asch, 1948; Fiske & Rai, 2015; Janis, 1983). Furthermore, the instinctual processes do not explicitly signal trustworthiness but instead rely on the assumption that people are more cooperative with group members than outsiders (B. Hare & Woods, 2021; Redish, 2022; Schelling, 2006; Sober & Wilson, 1998; Tomasello, 2016; Turchin, 2018).

These signals are driven by repeated experiences likely to imply group co-membership, such as taking meals together, experiencing hardships together, and sharing emotional states such as laughter or tears (Bastian et al., 2014; Boehm, 2012; Eisenberger, 2012; Gothard & Fuglevand, 2022; B. Hare & Woods, 2021; Marsh, 2017; Sander et al., 2003). In humans, group membership is often aligned with explicit identifications of identity, making narrative a key factor in community construction through instinctual processes (Boehm, 2012; Fiske & Rai, 2015; McCullough, 2020; Pinker, 2012; Redish, 2022; Turchin, 2018; D. S. Wilson, 2002). These instinctual processes evolved to recognize group membership through signals that were historically correlated with group membership of kith and kin, such as cultural and identity-based similarities.

Since community membership is transferable, particularly in humans (Boehm, 2012; McCullough, 2020; Milgram, 2017), instinctual trust should be transferable: if a member of the group vouches for a new member, that new member is likely to gain entry into the in-group level of instinctual trust. Conversely, mistrust can transfer to others in a group when violations arise. Within the group, if the violation of trust is an individual's misbehavior, then it might lead, instead, to rejection of the member from the group. This shows the importance of group rejection of individual misbehavior. Because transitions from in-group to out-group are consequential, instinctual trust is relatively insensitive to minor violations—but once a threshold is crossed, rejection (i.e.ostracism, ejection from the group) may be sudden and complete, even violent (Boehm, 2012; Ostrom, 1990; D. S. Wilson, 2002, 2015).

Instinctual trust arises from many small interactions that build a sense of community, which implies that small, positive, repeated interactions will create stronger bonds than a single community-building exercise. Moreover, because a community is defined by its members rather than by situational context, instinctual trust will depend more on the people one shares a situation with than the situation itself. Instinctual trust also rests on the assumption that community members are safer together — surviving intense shared experiences (e.g., combat, natural disasters) can produce particularly strong instinctual trust bonds (Bastian et al., 2014; Turchin, 2018). Knowing the mechanisms underlying instinctual trust enables us to recognize and use these strategies to increase trust, such as fostering diversity to create tolerance (Allport, 1979) and curating many small repeated interactions to produce camaraderie (Edmondson, 1999; Stickel, 2022).

## Procedural Trust

*Procedural trust* arises from habituated cooperative actions. The Procedural System stores disciplinary knowledge that enables rapid recognition of situations — social and otherwise — from complex perceptual cues (Ericsson et al., 2018; Klein, 1999) and releases corresponding action chains (Dezfouli & Balleine, 2012; Graybiel, 2008; Johnson et al., 2007; Redish et al., 2007). Both the action and the cues to release that action must be learned, but because they are learned, the set of available actions is not limited as is the case for the Instinctual system. Computationally, procedural action chains are released by the perceptual recognition of regularities in the environment. These regularities are learned gradually (Ericsson et al., 2018; O'Keefe & Nadel, 1978; Redish, 1999, 2013), are often not explicitly identifiable (Klein, 1999), and arise from well-understood neural processes that complete neural activity patterns from partial representations (Hebb, 1949; Hertz et al., 1991). Development of those action chains occurs through repetitive practice. Expectations of the future are implicit within these automated action chains, and the Procedural System does not store or process any information about why things happen. Rather than causal information or predictive information, it extracts and stores sequential information as its resources. Specifically, it stores the sequence of actions within a practiced action chain and the sequence of action chains to release in familiar situations. See Figure 3.

Figure 3: Procedural decision making.

KEY FACTORS:

- Information is represented in sequential action chains.
- Expectations are implicit due to recognition of situation patterns and release of action chains.
- Releases well-practiced action sequences in familiar situations.

COMPUTATION: The procedural system learns action chains. Given a complex sequence of decisions (top), if the same sequence is practiced in the same conditions, and is reinforced with positive feedback, it eventually turns into an action chain (bottom), such that the initial conditions can release a sequence of actions, based on properties of that initial state, generally precise but inflexible in its actions, and resistant to changes in the outcome because the procedural system does not explicitly represent the consequences of the action chain.

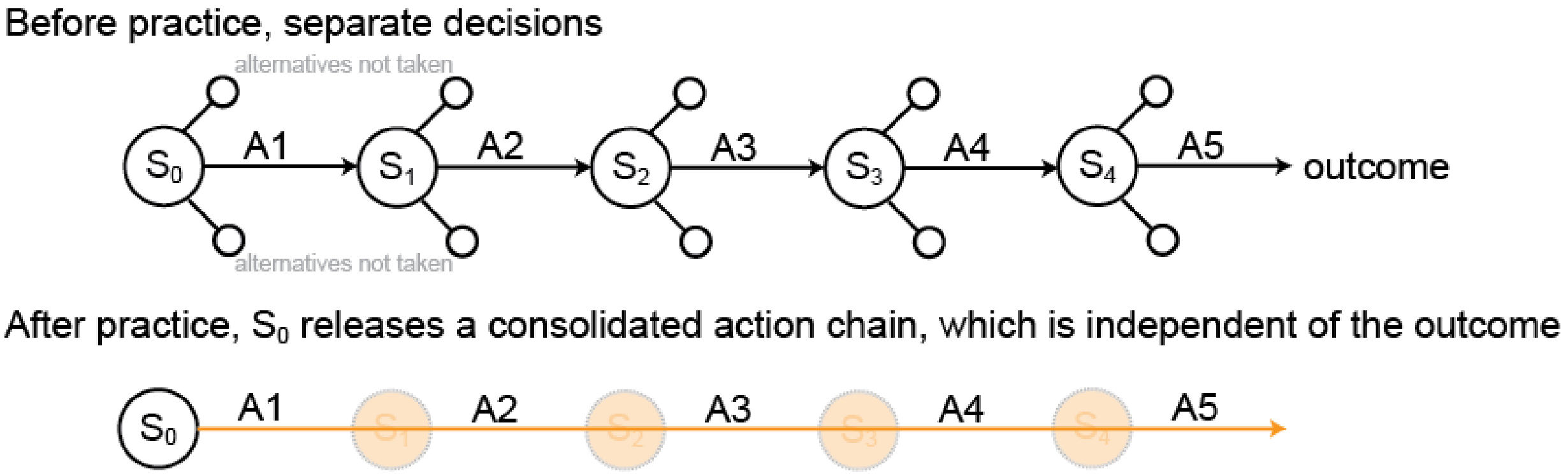

Neurophysiologically, the procedural system arises through circuits in the basal ganglia, a key component of which is the striatum, particularly the dorsal and lateral striatum (identified in primates as the dorsal caudate and lateral putamen) (Albin et al., 1989; Graybiel, 2008; Mugan, Amemiya, et al., 2024; Redish, 2013; Yin et al., 2004). The dorsal striatum receives input from cortical circuits and it is this corticostriatal connection that is hypothesized to learn to take the procedural actions, primarily through repeated reinforcement (Fisher et al., 2017; Houk et al., 1995; Jog et al., 1999; Santos et al., 2015). In addition, a key component of the procedural system is the primary motor cortex and its associated premotor and supplementary motor areas (Cisek, 2007a, 2007b; Tanji, 2001). Current neuroscience theory suggests that these cortical motor systems learn neural pattern sequences that can be modeled as high-dimensional paths through representational manifolds that play out motor behaviors learned in response to reliable experiences (Afshar et al., 2011; M. M. Churchland et al., 2006; Hatsopoulos et al., 2003). While there is evidence that cerebellum also plays an important role in timing and the actual execution of these sequences (Miall, 1998; Seidler et al., 2002), most models do not suggest a role for cerebellum in action-selection (Doya, 2000; Ito, 2000; Kawato, 1999; Llinás & Welsh, 1993). The data more strongly suggests action-selection arises through an interaction between the basal ganglia and the motor cortex.

*Procedural trust* exists when the environment is sufficiently reliable that one can respond to a given situation without explicitly considering the consequences of one's actions or making assumptions about social structure. For example, in a well-honed medical surgical team or in a well-practiced sports team, members can safely assume that the others will reliably perform their roles. A surgeon trusts the nurse to hand them the right tool at the right time and the anesthesiologist to monitor the patient appropriately. A football player trusts that their teammate will run the appropriate route to be at the right place to receive the ball. Likewise, ordinary social situations entail many habitual social actions.

The inflexible nature of Procedural resources requires procedural trust to be relearned when team members change, unless strong institutional standardization ensures consistent behavioral regularities. At its core, procedural trust is about regularities in the environment: if teammates are reliable, one can trust that they will be in the right place at the right time to complete the task.

Most experimental studies of trust are not based on the kinds of fast paced and reliable physical interactions that would be expected to primarily drive procedural trust; however, striatal signals associated with trust do show up in repeated interactions (Bellucci et al., 2020; Fouragnan et al., 2013), particularly when behaviors are well fit by reinforcement learning models (Fouragnan et al., 2013).

## Other computational components

In addition to the action-selection circuits described above, there are other neurophysiological components that are critical to prosocial behaviors, particularly the temporo-parietal junction (TPJ) and the ventromedial prefrontal cortex, which are involved in the identification and valuation of social constructs (Brethel-Haurwitz et al., 2022; Hutcherson et al., 2015; Lockwood et al., 2016). However, how these components interact with the primary decision systems is still a subject of much current research. Nevertheless, we can identify several additional computational components that are fundamental to making decisions. The extent to which these additional components are separated within decision systems or are cross-decision computational components is still debated.

*Perception* is a process that turns sensory signals into situational representations that can be used by the action-selection systems. Neurophysiologically, perception entails encoding information about the world in neural manifolds in sensory cortical systems, with increasingly abstract categorizations along cortical stages (Freedman et al., 2001, 2003). Most likely, TPJ is a sensory system dedicated to categorizing information about social situations. Whether these categorizations ever become explicit or remain implicit through distributed representations is still debated. Nevertheless, these computational processes underlying perception create consequences such as framing and priming (Kahana, 2020; McClelland & Rogers, 2003; Runge et al., 2023; Schacter & Buckner, 1998). Learning processes create expertise in which one learns to attend to the important dimensions for a given situation (Ericsson et al., 2018; Klein, 1999; McClelland & Rogers, 2003).

*Motivation* is central to deliberative evaluation and to learning in both the Procedural and Instinctual systems. Although specific aspects of motivation are still debated (Balleine, 2004; Niv et al., 2006; Redish, 2013; Sharpe et al., 2021), neurophysiologically it involves the nucleus accumbens core and shell, the orbitofrontal cortex, and the ventromedial prefrontal cortex (Ambroggi et al., 2011; Bartra et al., 2013; Chib et al., 2009; Padoa-Schioppa & Assad, 2006). There is important evidence that additional motivation information (such as computations that take longer to evaluate, like that one should choose healthy foods over sweet foods) require additional inputs into these areas (T. A. Hare et al., 2009; Knutson et al., 2007; Rangel & Hare, 2010). Whether motivation is separable across decision systems or whether there are central motivation components that maintain representations of value across all decision processes is still debated (Bartra et al., 2013; Chastain & Redish, 2025; Levy & Glimcher, 2012; Rangel & Clithero, 2014), but some computational processes of motivation are well-established. Most importantly, much of the underlying basal structure of motivation is based on intrinsic goals that are fundamental to a species (Edelman, 2014; Redish, 2013; Uchibe & Doya, 2008). There is evidence that cooperation and inequity aversion, particularly in the form of parochial altruism, is an intrinsic goal of human behavior (Fehr & Schmidt, 1999; B. Hare & Woods, 2021; Redish, 2022; Seymour et al., 2007; Singer et al., 2006; Tomasello, 2016). However, motivation itself is also subject to learning: for example, tastes can change after experiences such as chemotherapy (Bernstein, 1978; Jacobsen et al., 1993). Not surprisingly, reward-related signals often show up in social and neuroeconomic experiments, often with an increased reward arising from social closeness (Fareri et al., 2012; Hutcherson et al., 2015; F. Krueger & Meyer-Lindenberg, 2019; Lockwood, 2016; Lockwood et al., 2016; Marsh, 2017).

In short, the action-selection systems are not the only components of the decision process with learned resources. Research on trust should account for these perceptual and motivational learning effects, but since trust enables cooperation as an action, our taxonomy of trust starts from the resource differences in the three action-selection systems.

## The opposite of trust

If trust leads to taking actions that could make oneself vulnerable, that one takes because one believes that the actual risk of loss is low and the likelihood of cooperation is high, then there are multiple constructs that can serve as the opposite of trust (Bigley & Pearce, 1998; Cook et al., 2007; Ellickson, 1994; Lewicki & Bunker, 1995; Luhmann, 2018). One key construct is *distrust*, which can be phrased in the language used here as information that the action is likely to be more risky, that the likelihood of cheating and defection is high, making the actual risk of loss high. Another could be *indifference*, in which one takes actions independent of the choices made by the others in one's community. Finally, there could be *ignorance, uncertainty, or ambiguity*, in which one does not know what the trustworthiness of the outcome is. Our taxonomy actually provides for different answers to this question as a function of which decision-making system is driving behavior.

For Instinctual systems, the opposite of "trust" is almost certainly "distrust". Because the trust information processing step in instinctual systems is explicitly identified as the social identification of group behavior (Brewer, 2008; Dawes, 1980; De Dreu & Kret, 2016), the opposite of instinctual trust is most likely to be an identification of the individual as part of the outgroup, which is distrusted, often violently (Brewer, 2001; Diamond, 2013; Fiske & Rai, 2015). Many early biological attempts to address moral questions assumed that humans are parochial altruists, cooperating within communities, but xenophobic outside of them (Turchin, 2018; D. S. Wilson, 2015; Wright, 1995); however, more nuanced modern assessments of these perspectives argue that only the instinctual system is consistently parochially altruistic, while the other decision-making systems provide for social constructs that move beyond simple ingroup/outgroup dynamics (Greene, 2013; McCullough, 2020; Pinker, 2012; Redish, 2022).

In contrast, for Deliberative systems, “trust” arises from an explicit expectation of the other’s responses to one’s actions, and thus one can “distrust” deliberatively (I know you’re going to cheat. You always do.), but one can also be “indifferent” or even “ignorant” of the likelihood of betrayal. An agent would be indifferent to the outcome predicted if there were remediation mechanisms (such as insurance) that can ensure safer outcomes even if one were to be betrayed. Cook, Hardin, and Levy (2007) make the case that one can cooperate without trust through such remediation mechanisms, by defining trust only as a statement about the character of the partner, which leaves the opposite of trust to include an “indifference” perspective. However, these remediation mechanisms (such as insurance premiums) often impose a transactional cost that could be reduced if one were able to shift behavior into a more trusting condition that would not require paying for that mediation (Ellickson, 1994; Ostrom et al., 1992; Redish, 2022; Ullmann-Margalit, 1977). An “indifferent” antonym to trust does not remove trust from the decision process, but rather provides an opportunity to side-step it, potentially at a cost. Deliberation, however, also includes a “distrust” antonym to trust based on explicit expectations of betrayal. Gambetta and Hammill (2005), for example, cite signals that taxi drivers used to identify both decreases in risk (trust) and increases in risk (distrust). Including both ignorance and distrust in our understanding of deliberative antonyms for trust provides a clear explanation for the progressive risk taking seen in many new social encounters (see ***Building Trust*** below).

Because Procedural systems are learned sequences that only appear when the environment (including the social environment created by conspecifics) is reliable, procedural actions only arise when the social world is stable. They do not stabilize if the consequences of one’s actions are not reliable and a positive return on those actions are not consistently observed. Thus, procedural distrust also entails a similar process of stabilized action chains. Certainly, non-cooperative procedural behaviors can arise, as in the person who learns to do things on their own because cooperative behaviors have never worked out for them. Similarly, people learn to take procedural actions that increase safety in unsafe environments (such as learning defensive driving techniques). But these procedural behaviors also arise through positive reinforcement of those actions. Thus procedural actions reflect the stable expectations of that environment, which can engender cooperation (trust) or individualism (distrust).

An important question is whether trust and distrust can coexist simultaneously or whether they simply cancel each other out in a linear manner. The multifarious nature of trust implies that trust is not unidimensional, and thus suggests that it should be possible to manifest both trust and distrust at the same time — in a sense, this can be seen as motivational conflict. Motivational conflict is well-studied within the decision-making literature (Calvin et al., 2025; Capuzzo & Floresco, 2020; Damphousse, Calvin, et al., 2026; McNally, 2021; Redish, 2016) and is hypothesized to be a key component underlying anxiety (Beck et al., 2005; Gray & McNaughton, 2000; Mobbs & Kim, 2015). These conflicts can certainly happen between decision systems, wherein one system trusts and another does not. There are many cases of people trusting instinctually (because of familial kin relationships) while knowing that the person is likely to betray them (deliberative distrust). Alternately, one sees "white knuckle cooperation", where one deliberatively trusts but instinctually does not. But one can also see trust and distrust in conflict within systems. For example, the deliberative system runs scenarios forward; if those scenarios sometimes lead to trust and sometimes to distrust, the difficulty of resolving these scenarios can lead to dithering and deliberation. Because humans have complex group dynamics, people can be members of multiple groups, and instinctual in-group definitions can compete with each other, leading to a conflict between trust and distrust. Given the learning processes in procedural trust, we predict that motivational conflict within the procedural system alone will be unlikely, but it is certainly possible for procedural action probabilities to conflict with instinctual or deliberative decisions. As with other forms of motivational conflict, we suggest that unresolved conflicts between trust and distrust are likely to lead to psychological distress, discomfort, and anxiety.

## Neurophysiological studies of trust and distrust

We can test our multiple-decision-system hypothesis of trust by looking at the cognitive neuroscience literature applied to making decisions in the face of social risk. For simplicity, we will restrict this test of our multiple-decision-system hypothesis to the human literature as trust interactions between conspecifics depend fundamentally on the ethology of the species, and we do not want to assume that this ethology is maintained across species. There are, of course, lots of experimental data looking at trust behaviors across species, particularly in other primates (de Waal, 1990; Goodall, 1986), but also in bats (Carter et al., 2020), birds (Emery et al., 2004; Raby et al., 2007) and rodents (Ben-Ami Bartal et al., 2011; Sun et al., 2025). While these cross-species questions are fascinating, we will not address them in this manuscript due to limitations of space.

We can make four specific predictions from this framework:

1. The multiple neural systems theory predicts that different neural circuits will be activated in different decisions, but that these neural systems will be consistent within themselves. Thus, deliberative decisions should arise from a specific network based on prefrontal cortex, hippocampus, and orbitofrontal cortex, while instinctual decisions should depend on amygdala, anterior insula, and the subgenual anterior cingulate cortex, and procedural decisions should depend on motor cortex and the dorsolateral striatum.

2. Different trust processes should activate different neural circuits. Trust decisions made by ingroup/outgroup considerations should activate instinctual circuits more than procedural or deliberative, while trust decisions made logically through explicit expectations and remediative consequences should activate deliberative circuits. Trust decisions that arise procedurally should activate dorsal striatal circuits.

3. Which neural system is activated should be more predictive of decision choices than the situation one is in. Because any given behavior can arise from multiple neural causes and because multiple neural systems can be drawn into use in any given situation (Constantinople, 2026; Redish et al., 2022), neural circuit activation should be more predictive of outcome than the scenario itself. Importantly, scenarios are likely to be correlated with neural circuits, which can lead to different choices in different scenarios, but the neural circuit should be the key difference between outcomes.

4. The activated neural system should predict the subtleties of decision choice. Instinctual trust should follow from in-group/out-group recognition, deliberative trust from explicit expectations, and procedural trust should only arise in well-practiced conditions.

One of the first places that these differences were seen is in the trolley problem (Greene et al., 2001; Moll et al., 2001). In the trolley problem, the subject is asked to choose between taking an action that kills one person and not taking the action, which kills five. Differences are seen depending on the proximity of the action to the death itself, as well as various other manipulations that are likely to shift between instinctual and deliberative systems (Costa et al., 2014; Edmonds, 2015; Greene, 2013). While these complex scenarios create correlative shifts between being willing to take the action or not, there is a stronger relationship to the neural circuits involved — taking the deontological inaction of refusing to kill vs. the consequentialist action of choosing five lives over one is directly related to neural activity in the amygdala (instinctual system) vs. neural activity in the dorsolateral prefrontal cortex (deliberative) (Greene, 2013; Greene et al., 2001).

A similar distinction is seen in the prosocial literature that differentiates consideration of others as “empathy” versus “mentalizing” (Bellucci et al., 2020; Healey & Grossman, 2018; Sandhu et al., 2026; Thai & Lockwood, 2024; Tusche & Bas, 2021). Empathy entails an emotional imagination (and recapitulation) of the other’s pain or joy (Batson, 2016; Lockwood, 2016; Singer, 2008), while mentalizing entails a prediction of what someone will do, understanding their perspective and their decision process (Frith & Frith, 2006; Knudsen & Wallis, 2022; Premack & Woodruff, 1978; Tusche & Bas, 2021). In discussions of cooperation and competition, empathy is more limited in that it can only drive behaviors that match the other’s behavior (cooperation with cooperative behaviors, refusal in the face of defection), but mentalizing can be used more generally to take actions based on the expectation of outcome, whether the partner’s behavior is matched cooperatively or competitively or not (Redish, 2022). In the three-system language used above, empathy is an instinctual process, while mentalizing is a deliberative process. Importantly, because deliberation is more cognitively taxing and is often cut short due to working memory limitations (J. R. Anderson & Schooler, 1991; Kahneman, 2011; Kurth-Nelson et al., 2012), empathy can often find a cooperative solution when simple deliberative expectations lead to pathological individualist decisions (Redish, 2022).

Neurophysiologically, the hypothesis that empathy arises through instinctual systems and mentalizing through deliberative tracks the observed neural activity. An extensive literature has found that empathy activates salience circuits (amygdala, anterior insula, subgenual anterior cingulate cortex), while mentalizing activates the default mode and prefrontal executive networks (Bellucci et al., 2020; Gu et al., 2013; Sandhu et al., 2026; Thai & Lockwood, 2024; Tusche & Bas, 2021). In humans, thinking about a friend's pain (ingroup) activates the anterior insula and subgenual anterior cingulate cortex while thinking about a stranger's pain (outgroup) activates dorsolateral prefrontal areas, suggesting that ingroup empathy is more instinctual, while outgroup recognition of pain (mentalizing, perhaps definable as "sympathy") is more deliberative (Eisenberger, 2012; Sandhu et al., 2026; Tusche & Bas, 2021). In fact, anterior insula is a critical component in empathy, as people with anterior insula lesions show diminished empathy responses (Gu et al., 2012; Singer, 2008). Similar differences are seen in amygdala activity, showing more activation for more altruistic individuals who see the larger community as ingroup, and less for more sociopathic individuals, who often see themselves as members of a community of one (Marsh, 2017).

The neural underpinnings of trust have been studied in experimental behavioral economic games (Rățală & Sanfey, 2026), including the ultimatum and dictator games (Güth et al., 1982; Güth & Huck, 1997), the investment or trust game (Bellucci et al., 2017; Berg et al., 1995; King-Casas et al., 2005), and the public goods or tax game (Fehr & Fischbacher, 2003; Marwell & Ames, 1979). In the *ultimatum game*, one player divides money between themself and another, and the other can either accept or reject the deal. In the *dictator game*, the second player is not given a choice, and the money is split as the first player decides. In the *investment game* (sometimes called the *trust game*), one player moves some amount of money to another, which increases in the transfer — the other player can then return some of the money or not. In the *public goods game*, a community of players invests money in a central pot, which increases in the transfer, but then pays out equally to all players, independent of the amount invested by the individual. Both one-shot and repeated-play versions of these games have been studied. These games provide an experimental space with which to explore the neurophysiology of trust and cooperation. As our goal in this paper is to address questions of trust specifically and not the general question of prosociality or community, we are not going to address these more general questions, but point readers to large scale reviews of these games (Camerer, 2011; Fehr & Fischbacher, 2003; Glimcher et al., 2008; Rățală & Sanfey, 2026; Redish, 2022; D. V. Smith et al., 2026), to large scale reviews on the neurophysiology of prosociality (Eisenberger, 2012; Greene, 2013; Redish, 2022; Sandhu et al., 2026; Thai & Lockwood, 2024), and to large scale reviews on prosociality itself (de Waal, 1990; B. Hare & Woods, 2021; McCullough, 2020; Ostrom, 1990; Pinker, 2012; Redish, 2022; Sober & Wilson, 1998; D. S. Wilson, 2015).

Neurophysiologically, emotional rejection of unfair responses in the ultimatum and investment games activates instinctual systems (such as anterior insula) (Aimone et al., 2014; Bellucci et al., 2017; Sanfey et al., 2003) and implicit recognition of social signals within amygdala (Engell et al., 2007). In the investment or trust game, explicit recognition of the history of reciprocation activates deliberative systems such as prefrontal circuits while slowly learned reputation recognition activates procedural systems such as striatal circuits, increasing with cooperation, but not betrayals (Bellucci et al., 2017; Fouragnan et al., 2013; King-Casas et al., 2005). In a fascinating experiment, Hughes et al. (2017) specifically examined trust behaviors within group vs. between groups. They found that ingroup trusting behaviors could arise from either prefrontal activations (primarily involving deliberative structures) or from ventral striatal (accumbens, involving instinctual structures). They found that ingroup trust could arise from either system, but outgroup trust only appeared with time and deliberative system activation. King-Casas et al. (2008) found that people with borderline personality disorder (who have particular difficulties with trust) had diminished signaling in the anterior insula in the investment game, which likely prevented them from coaxing trusting behaviors from partners over repeated cycles. In a variant of the ultimatum game designed to test for spite in which players decide whether to give another player the change to share with them or to pay a cost to punish them (to spite them), persecutory ideation (paranoia) was correlated to salience network activation, but rational mistrust (when the other player gained more by punishing them) was correlated with lateral frontal activation (Kazinka et al., 2024; Wisner et al., 2021).

These experimental data strongly support the idea that these separate neural systems are driving different pro- and anti-social behaviors under different scenarios. Our review above identifies that these separate neural decision systems process information differently, are activated in different social scenarios, and come to different conclusions, leading to different behaviors. If we define trust as the step in the information process that makes one willing to make oneself vulnerable and to take a physical, financial, emotional, or social risk, then each of these systems will use different information to make that assessment that it is safe to take the risk, and they will lead to different kinds of trust.

This analysis has important consequences for social construction. Changing the social environment will change both which system is most likely activated to drive behavior and how that system assesses the risk. Thus, changing the social environment (through individual and institutional choices) will change the willingness to trust.

## Building Trust

Trust is fundamentally a question of coordination, modeled in Game Theory as a *coordination or assurance game* (Binmore, 2005; Dawes, 1980; Fehr & Fischbacher, 2003; Kollock, 1998; Marwell & Ames, 1979; Redish, 2022; Runge, 1984; Skyrms, 2004; Ullmann-Margalit, 1977). In these games, if one's partner is going to cooperate, then the optimal choice (the best option for the individual making the choice) is to also cooperate, while if one's partner is going to compete (i.e. "cheat" or "defect" in Game Theory parlance), then the optimal choice is to also compete. This interdependence produces two stable equilibria: groups of cooperators and groups of competitors (Kollock, 1998; Redish, 2022; Runge, 1984; Ullmann-Margalit, 1977). Empirical studies confirm this logic: humans generally cooperate when others cooperate and compete when others compete (Fehr & Fischbacher, 2003; Marwell & Ames, 1981). These dynamics lead to virtuous cycles, where cooperation begets cooperation, and vicious cycles, where competition fuels further competition (Axelrod & Hamilton, 1981; Redish, 2022; D. S. Wilson, 2015). Crucially, systemic conditions shape which cycle dominates. Small changes in institutional design, incentives, or shared narratives can shift groups toward sustained cooperation or entrenched conflict (Fehr & Fischbacher, 2003; Ostrom, 1990; Pinker, 2012; Redish, 2022; Ullmann-Margalit, 1977; D. S. Wilson, 2015). Understanding how trust is built, violated, and repaired therefore requires examining the learning mechanisms and information representations within each decision system.

## Progressive Risk Taking and Virtuous Cycles

People tend to cooperate when others do, but trust rarely develops in a single act because being vulnerable means risking losses or sacrificing potential gains. To balance self-preservation, information gathering, and social signaling, individuals often engage in progressive risk-taking: starting with small risks that limit losses if trust is misplaced, and gradually escalating to greater vulnerability. Progressive risk taking can be seen in many non-human animals as well, such as primates deciding whether to defend one-another in conflicts within and between tribes (de Waal, 1990; Goodall, 1986; Strum, 2001), or vampire bats deciding whether to share precarious food supplies gained each night with partners who may or may not reciprocate the next night (Carter et al., 2020). Partner reciprocity provides information about whether to proceed with increasing the vulnerability. Likewise in human partnerships, a good strategy is to step through a series of increasingly vulnerable actions, each of which provides information about how to safely proceed to the next (Binmore, 2005; Boehm, 2012; Cronk & Leech, 2012; Redish, 2022; Stickel, 2022; Ullmann-Margalit, 1977). Too much vulnerability too soon risks personal loss. Too little vulnerability restricts information gathering and may signal an unwillingness to cooperate. Along with considering the acceptable degree of vulnerability, accomplished practitioners in building trust strategize around 1) the specific risks each agent can accept, 2) what information is being gathered, and 3) what signals are being sent in each act of cooperation (Edmondson, 2018; Simons, 2008; Stickel, 2022). By explicitly modeling the information regularities involved in each decision-making system (deliberative, procedural, instinctual) one can predict which form of trust will arise from a given cycle of progressive risk taking. Causal information builds deliberative resources, repeated sequential actions builds procedural resources, and predictive signals of group membership and in-group safety build instinctual resources.

Since each act of cooperation may gather or provide information that is causal, sequential, or group-predictive, a virtuous cycle may build on multiple steps of deliberative, procedural, and/or instinctual trust. *Deliberative trust* could increase vulnerability through explicit steps, identifying specific risks, outcomes, and expectations. Observing a partner's behavior allows past actions to inform explicit predictions of future actions, building a mental map of expectations. As potential risks are resolved, the mental map will reflect the extent to which those expected consequences reflect the likelihood of the other agent cooperating. *Procedural trust*, in contrast, would likely depend on repetitive practice to automate essential routine procedures. New team members can practice important procedures with the team under close monitoring, and progress to less oversight over time. These sessions involve the vulnerability of time, effort, and potential embarrassment, but carry less risk than immediately entering high-stakes procedures. *Instinctual trust* can be built by initiating small social interactions, despite the affective discomfort when available instinctual signals do not yet indicate partnership or group belonging. Because the Instinctual System depends on social signals and context, instinctual trust arises most reliably when in-group cooperation is experienced across diverse situations. In the early stages of building instinctual trust, high levels of trust in one situation may not initially carry over to another situation.

The decision systems can build on each other. After developing deliberative trust through explicit expectations (perhaps with punishments for non-compliance), the resulting social structure can help create the reliable conditions needed for procedural trust to develop. Likewise, instinctual trust from shared identity or shared narratives can jump-start the development of deliberative trust. An important step in this process is information sharing. Sharing the reasons for one’s actions improves predictability and strengthens deliberative trust. Likewise, finding common ground and shared identities can create shared connections to create an in-group, which would increase instinctual trust.

## Viscous Cycles, Forgiveness, and Error-Correction

People tend to compete when others do, and thus trust can collapse even among legitimately trustworthy partners when errors trigger a downward spiral. Such spirals can be repaired through forgiveness and reconciliation (R. A. Anderson et al., 2020; Axelrod, 1986; Crockett et al., 2017; Gesiarz & Crockett, 2015; Ostrom, 1990; Redish, 2022; D. S. Wilson, 2015). Our taxonomy of trust predicts that each form of trust has unique error modes arising from the limitations of each decision system, and that breakdowns in each kind of trust can be recovered differently by accessing the information processing in that decision system.

Since the Deliberative System stores information about how and why events occur, rebuilding deliberative trust requires examining the underlying cause of the error. This includes reflecting on the rationale behind one’s own actions. Because behavior reflects the interaction of all three systems, people do not always have immediate insight into why they acted differently from how they predicted they would (Eagleman, 2011; Gazzaniga, 2011; Kurzban, 2010). Deliberative trust allows people to act cooperatively in predictable situations, even new situations (if they have predictable consequences), but it is not usually the main driver of actions in routine situations where the Procedural System can dominate or in entirely unfamiliar situations where the Instinctual System can dominate. Current theories of decision-making predict that behavior is often driven by the parallel, complementary interaction of all three systems. One error mode for deliberative trust is that people may enter situations expecting to cooperate but find themselves competing due to instinctual or habitual responses they had not anticipated. In such cases, rebuilding deliberative trust requires not just reflection but also situational changes that shift decisions to deliberative processes.

Since the Procedural System stores patterns to recognize and actions to release, it does not store information about how and why things happen. We can predict that one error mode for procedural trust is rigidity in performing previously learned actions (“force of habit”). A second error mode is the failure of procedural resources to transfer to new situations. Responding to procedural errors solely with deliberative or instinctual interventions is likely to be ineffective. Discussing the cause of the error, providing team members with assurances about group membership, and addressing any negative feelings arising from the error are good practices to support deliberative and instinctual trust, but do not build procedural trust. Procedural trust is not restored by identifying errors but only through renewed practice. Success in domains that rely on procedural actions depends on persistence: getting back on the horse after falling, retrying a failed operation, relearning the components of the individual action-chain (Ericsson et al., 2018).

Instinctual trust requires repeated positive social interactions and thus requires social glue to recover from trust violations (Boehm, 2012; B. Hare & Woods, 2021; Redish, 2022; D. S. Wilson, 2002). One error mode is the disparity in the timescale and situational sensitivity between learning new group identities and signaling existing group identities. Developing new instinctual trust is slow, while signaling existing instinctual trust is fast. As a result, people may remain strongly influenced by past instinctual resources, even when these conflict with current goals or values. Just as someone may relapse into substance use despite deliberatively seeking sobriety, people may maintain deep affection for a former partner after a breakup, or retain long-term prejudices despite explicit changes in beliefs. People can also maintain animosity after a past betrayal has been corrected. Importantly, these phenomena are consistent with the type of information that the instinctual system extracts from the environment and the probabilistic calculations it performs. Thus, error correction of instinctual resources requires the appropriate information regularities. An apology, for example, is a marker that the person who violated the trust recognizes it and is going to (theoretically) learn from it. When people find themselves in a situation of greater vulnerability than their instinctual resources support, their instinctual systems may release a fearful somatic state. If others act cooperatively in such high-vulnerability, high-arousal states, the resulting relief can teach the instinctual system to expect cooperation in similar situations.

Importantly, it is also possible to use one form of trust to help repair a breakdown in another. For example, a deliberative process such as a legal sanction can temporarily enforce cooperation during an instinctual breakdown of trust, creating time for instinctual trust to be rebuilt. An appeal to an instinctual community can be used to rebuild a lack of trust in the deliberative system. The repetition of a procedural routine to create a ritual can be used to rebuild a feeling of community and help restore instinctual trust.

## Institutional and Organizational Effects on Decision Systems Impact Trust

Institutions and organizations can empower or erode trust by complementing or interfering with human decision-making processes. Institutions and organizations provide the regularities in the environment that the decision-making systems extract and store as resources. Institutions and organizations can also perform information processing operations providing additional informational resources to human decision-making systems. Certifications, compliance checks, and other explicit markers can provide an opportunity to transfer trust from one agent to another. Reputational markers can identify people to be safe to trust and those not to in larger societies. Scientific reports that report all methods, including flaws, and those that include the raw data provide increased opportunities to trust the results. Thorough documentation can provide evidence of a system's intended steps to take. Similarly, demonstrated punishments for breaking trust and non-compliance can be a signal that an institution will not tolerate trust-breaking, while lack of punishment, non-transparent investigations, and lack of consequences for non-compliance lead to a lack of trust in institutions. The societal institutions we have created have strong effects on trust and are an actionable avenue to build trust (Cook et al., 2007; Cronk & Leech, 2012; Ellickson, 1994; Gambetta & Hamill, 2005; McCullough, 2020; Ostrom, 1990; Pinker, 2012; Redish, 2022; Stout, 2010; Ullmann-Margalit, 1977; D. S. Wilson, 2002). We propose that these societal institutions can more reliably build trust and repair violations of trust when training practices and accountability policies strategically account for the fundamental information regularities inherent to each form of trust.

While the general observation that institutions can affect behavioral manifestations of trust is well accepted, how those effects arise depends strongly on the mechanisms through which they interact with trust. Behavior is fundamentally an interaction between the external environment and the internal information processing within the brain (Redish, 2013), which means that one can effect changes either by changing the internal processing or by changing the external environment (Redish, 2022; Redish et al., 2025). Defining trust in terms of the cognitive neuroscience as we have done here suggests that institutional effects will interact with each of the decision-making systems differently to create different effects on trust. In fact, we would argue that those institutional mechanisms that do work to modulate trust, do so specifically because they access these fundamental cognitive neuroscience mechanisms.

Instinctual trust depends on the definitions of who is in the ingroup and who is not (Brewer, 2001, 2008), thus institutional mechanisms that define groups can modulate instinctual trust. Long term experiences within communities create community goals that go beyond the individual — for example, common resources are often well-protected within long-term communities, particularly those with strong community identification (Acheson, 2012; Allen, 1982; Ostrom, 1990; Runge, 1984). Many researchers have argued that religious identifications, nationalism, and other identity groups (sports team and other fan bases, shared corporate, school, or university identities, as well as clubs and other organizations) are all mechanisms to create community (Atran, 2002; Pinker, 2012; D. S. Wilson, 2002, 2015). Synchronized behaviors such as rituals serve to bind individuals together in a recognized group (Hobson et al., 2017, 2018; Pinker, 2012; Redish, 2022; D. S. Wilson, 2002). Several researchers have argued that difficult institutionalized entrance rituals increase the value of group identity (Aronson & Mills, 1959; Gerard & Mathewson, 1966; Sosis & Bressler, 2003) because of fundamental processes through which value judgements depend on the cost spent (Inzlicht et al., 2018; Kacelnik & Marsh, 2002; Mcafee et al., 2010; Plassmann et al., 2008; Sweis et al., 2018). All of these processes affect neural structures critical to the instinctual circuit, such as amygdala and the anterior insula / subgenual anterior cingulate cortical salience network (Eisenberger, 2012; Gothard & Fuglevand, 2022; Marsh, 2017).

Institutional mechanisms can impact the information processing of deliberative decision making in three ways that could affect trust. First, institutional mechanisms can change the expected consequences of behavior, such as explicit punishments for defection and betrayal (Cook et al., 2007; Fehr & Fischbacher, 2003; Ostrom et al., 1992; Rockenbach & Milinski, 2006). Second, institutions can provide for remediation of the consequences of betrayal, such as coverage in the face of loss (for example through insurance) or through explicit restorative justice to replace losses (Braithwaite, 2004; Gavrielides, 2005; Redish, 2022). Third, institutional mechanisms can serve as a reputation marker, making people more willing to change their expectations of the likelihood of betrayal (Alexander, 2010; Berg et al., 1995; King-Casas et al., 2005; Redish, 2022). Importantly, these institutional markers do not necessarily need to be governmental — rideshare companies purport to provide reputational markers for both drivers and passengers, reducing the complex decision processes that used to be necessary for individual taxi drivers (Gambetta & Hamill, 2005).

The key to procedural trust is coordination and the regularity of the behavior of one's conspecifics. Institutional codifications of rules and playbooks (such as which side of the street to drive on or the color of stoplights and stop signs) can provide for simplified decision coordination by regularizing that behavior (Redish, 2022; Ullmann-Margalit, 1977). Community rituals can also create procedural regularities that can simplify trust, such as when the members of a Grameen Bank collective meet weekly to pay their interest on their loans (Bornstein, 1999; Yunus, 2007).

Moreover, institutional choices can shift decisions between systems by reframing the question. Famously, changes in the framing of the trolley problem changes the decision system driving behavior (revealed in changes in the activated neural circuit), producing changes in the decision itself (Greene, 2013; Greene et al., 2001). This means that in addition to being a source of additional information processing mechanisms, external systems can also affect information storage through manipulations of environmental demands and resources. External systems are not constrained by the learning mechanisms or processing limits of the physical brain (Clark, 2008; Norman, 2013). Even a simple tool like a checklist can serve as an external system which overcomes the limits of working memory, simplifying deliberative processing, and regularizing behavior to increase procedural learning.

Organizations can also turn errors into opportunities to build trust. Amy Edmondson's research on work teams in hospitals has shown that serious patient care mistakes serve as a valuable opportunity for organizational learning (Edmondson, 1999). When hospitals failed to learn from their mistakes, it was often attributable to lack of trust within the system. Following the Challenger disaster, the engineering community made substantial changes to its accountability protocols (McDonald, 2012; Rogers et al., 1986; Werhane, 1991). These changes addressed not only technical reliability but also cooperation and accountability. Thus, at the organizational level, failures in performance can be used to improve performance (Edmondson & Lei, 2014), and failures in cooperation can be used to improve trust (Stickel, 2022).

We want to note, however, that these trust systems can also arise without institutional influence (Boehm, 2012; Ellickson, 1994; Redish, 2022). For example, procedural trust can arise naturally without codification. We all know couples that have a morning routine that they do every day for decades, or chores that are split, never explicitly negotiated, never codified, but nevertheless, reliable and trustworthy. Similarly, the patron who comes into their "third space bar" (Oldenburg, 2023) every day and the bartender who puts down their usual drink even before they sit down (or perhaps even has it prepared before they walk into the bar) are both showing examples of procedural trust without codification in rules and playbooks. Instinctual trust based on group dynamics has a long evolutionary history (Boehm, 2012; de Waal, 1990; Goodall, 1986; B. Hare

& Woods, 2021; Tomasello, 2016), and deliberative trust can arise based on specific recognition of individual actions (Bellucci et al., 2017; Berg et al., 1995; King-Casas et al., 2005).

## A comparison to other taxonomies of trust

Historically, there has been a large discussion about what trust is and trust is not. Several researchers create taxonomies by defining trust in a limited way and noting that one can get to cooperation without their definition of trust (Cook et al., 2007; Ellickson, 1994; Gambetta, 1988), while others have focused on specific aspects of trust, such as ingroup/outgroup effects (Brewer, 2001, 2008; De Dreu & Kret, 2016; Scheepers & Derks, 2016), while others have identified trust only in its counterpart to distrust and power (Luhmann, 2018).

One of the most influential taxonomies of trust was developed in the mid-1990s by Mayer, Davis, and Schoorman (1995), who proposed that trust is driven by “rational beliefs” about three trustworthiness subconstructs: *ability* (also called *competence*), *benevolence,* and *integrity*. Mayer et al.'s model (Mayer et al., 1995) assumes that learning occurs primarily through the trustor forming perceptions of these three attributes in the trustee, so errors in cooperation arise from misjudging traits. These dimensions have been widely applied in survey research, yet subsequent work has identified persistent psychometric problems. Studies find that these dimensions are not empirically distinct, that survey items lack conceptual clarity, and that observed separations are difficult to replicate (McEvily & Tortoriello, 2011).

Legood et al. (2023) argue that such weaknesses arise because these constructs of ability (competence), benevolence, and integrity are not truly orthogonal, making findings elusive and fragmented. They call for a shift towards antecedents rooted in well-understood cognitive and affective processes. Our approach addresses this call by arguing that vulnerability and cooperation depend on the information processing underlying three decision-making systems (deliberative, instinctual, and procedural). This framework explains why survey dimensions such as ability, benevolence, and integrity are unlikely to appear orthogonal — all three decision systems (deliberative, instinctual, procedural) include resources that represent and respond to information about ability, benevolence, and integrity. In contrast, our taxonomy grounds learning in the neuroscience of decision-making, which specifies distinct error modes and recovery pathways for deliberative, instinctual, and procedural trust. By focusing on the computations and learning processes specific to each system, our model yields concrete and testable predictions about how trust is built, violated, and repaired.

Another approach in the trust literature has been to build on dual-process psychological theories — whether fast vs. slow, heuristic vs. deliberative, affective vs. cognitive, impulsive vs. reflective, implicit vs. explicit (Healey & Grossman, 2018; Kahneman, 2011; McAllister, 1995; Murray et al., 2011; Strack & Deutsch, 2004; T. D. Wilson et al., 2000). Furthermore, while these dual-process models capture meaningful contrasts, they also assume that errors primarily stem from the heuristics of the ‘fast’ system or from failures of the ‘slow’ system to regulate it (Kahneman, 2011; Kahneman & Tversky, 1996; Kurzban, 2010), although some argue that heuristics can be efficient under conditions of bounded rationality (Delton et al., 2011; Gigerenzer & Brighton, 2009; Gigerenzer & Gaissmaier, 2011; Gigerenzer & Goldstein, 1996;

Lieder & Griffiths, 2019). In contrast, our framework emphasizes that all three decision systems contain both fast and slow components, each with unique error modes and repair mechanisms for trust. None is inherently more rational than the others and each is adaptive in different ways.

For example, although Kahneman's (2011) slow system (System II) captures some deliberative phenomena, it fails to account for how the Deliberative System learns, represents, and processes expectations of the future (Redish, 2016; Redish et al., 2025). For example, it does not account for the limitations of what is searched (Hunt et al., 2021; Huys et al., 2012; Redish & Johnson, 2007; Stewart et al., 2006), the complexity of the evaluation (Kahneman & Tversky, 2000; Redish, 2016; Shenhav, 2024), or the fact that deliberation can sometimes be rapid. The Deliberative System is not a "rational actor" in the economic sense; its reliance on available computations produces systematic inaccuracies (Redish et al., 2025; Stewart et al., 2006; Tversky & Kahneman, 1973). Evaluation is biased by immediate emotional states, a rejection of extreme conditions, asymmetric risk assessments in gains and losses, a dependence on the ability to bring past events into memory, and difficulty with probabilistic reasoning (Crockett et al., 2017; Kahneman & Tversky, 1996, 2000; Lichtenstein & Slovic, 2006; Runge et al., 2023; Simonson & Tversky, 1992; Tversky & Kahneman, 1973). Deliberation involves a depth-first search through categorized states of the world, tracing potential action-sequences serially in an imagined future that may or may not be the most likely outcome (Redish, 2016) This process depends on evaluating intermediate and future states (Huys et al., 2012; Kurth-Nelson et al., 2012), and its accuracy hinges on how concretely those states are represented (Peters & Büchel, 2010; Trope & Liberman, 2003). Finally, imagined outcomes are evaluated by neural circuits evolved for evaluating present conditions (Niv et al., 2006), making them sensitive to learned motivations and current needs (Andrade & Ariely, 2009). All of these processes can introduce inaccuracies that interact with one's environment, including one's social environment.

Other theories argue that these dual-process components interact to produce trust. For example, McAllister (1995) argues that trust contains both cognitive and affective components, where the cognitive components are based on explicit predictors of trust (which include, in his formulation, predictions of reliable performance, professional credentials, and cultural-ethnic similarity) and the affective components are based on experience such as frequent interactions. While superficially similar to our distinction between deliberative and instinctual trust, the overlap is limited, and we would argue that McAllister's components do not track with our distinctions directly.

We note that the Deliberative System is not restricted in identifications of any particular reason to cooperate such as transactional motives (or trustworthy traits). In fact, deliberative systems correctly predicting the other's competitive actions can enable one to trick the other into cooperation (Radiolab, 2014; Redish, 2022). Instinctual trust extends beyond emotional bonds to encompass the broader hereditary repertoire of innate behaviors and affective states. And, finally, procedural trust plays an important role in routine cooperation yet is absent from these dual-process models.

McAllister (1995) explicitly argues for a fixed cognitive-to-affective sequence in the development of trust. In contrast, we argue that cooperative trust arising from any one system can support the development of trust in any other. Rituals and cooperative behaviors such as practicing sports together, reciting a catechism, or marching in step can lead from procedural behaviors into instinctual trust; instinctual trust can build a community willing to accommodate legal systems that define deliberative trust.

Several neural models of trust are based on other computational subcomponents that can be mapped to our three-system taxonomy. Fareri (2019), for example, similarly starts from a definition of trust as social expectations that make one willing to assume risk, and notes that these signals can be explicit or implicit. While their definition of implicit expectations aligns well with our instinctual system, their definition of explicit signals includes both prior-informed signals (which we would identify as deliberative) and experienced signals (which could drive either deliberative or procedural trust in our taxonomy). We would predict that explicit recognition of betrayal and episodic recognition of positive and negative events would affect deliberative neural circuits, while slower changes to cooperative and defection events would affect procedural circuits.

Another neural model divides the computations necessary for trust into reward, salience, central-executive (strategy), and default-mode networks, suggesting that trust processes depend on using cognitive (reward expectation) and social (implicit expectation) signals to recognize treachery, trustworthiness, and a reduction in the uncertainty (F. Krueger & Meyer-Lindenberg, 2019). Like McAllister, Krueger and Meyer-Lindenberg suggest that trust always proceeds from explicit calculations to knowledge to identification. We argue, in contrast, that the three systems provide independent opportunities to reduce uncertainty and increase trust, and thus that one can create trust in any order — from observations of reciprocity (deliberative) to friendship (instinctual), but also from social group recognition (instinctual) to explicit calculus (deliberative), from deliberative and instinctual to experience of practice (procedural), but also from experience of practice (procedural) to instinctual and deliberative.

Other notable models parse trust by its referent — trust in an individual, an organization, or a technology (Fulmer & Gelfand, 2012; Hoff & Bashir, 2015; J. D. Lee & See, 2004; McKnight & Chervany, 2001; Rousseau et al., 1998; Zucker, 1986) or by stages of trust development (Hardin, 2004; Lewicki & Bunker, 1995; Rempel et al., 1985; Zucker, 1986). These approaches capture important patterns but along different dimensions than our neuroscience-based model. For example, when a frequent flyer goes through all the motions of packing for a flight, preceding through airport security, and boarding an airplane, our taxonomy classifies the trust inherent in these actions as a question of mechanism (deliberative, instinctual, or procedural trust) rather than as a function of whether the referent is the pilot, the airline, or the aircraft. Each system highlights different aspects of referents: deliberative trust depends on explicit expectations based on knowledge of individuals, training, or safety records; instinctual trust draws on contextual cues such as brand loyalty; and procedural trust emerges from repeated experience and is not referent-specific. Combining referent models with our framework could yield new insights into how decision systems interact with different trust targets.

## Prosocial constructs beyond trust

The concept of trust is only one component of the complex interaction that underlies human social structures (Boehm, 2012; Bowles & Gintis, 2013; Cook et al., 2007; McCullough, 2020; Pinker, 2012; Redish, 2022). It interacts with other complex societal constructs, such as fairness and inequity (Binmore, 2005; Fehr & Schmidt, 1999), shame and guilt and blame and forgiveness (R. A. Anderson et al., 2020; Katchadourian, 2011; Ostrom, 1990), as well as reciprocity expectations and indebtedness (Ellickson, 1994; Gao et al., 2024). Importantly, these complex constructs depend on assumptions about the motivation of the other agents in the community, particularly differentiating between intentional and accidental consequences (R. A. Anderson et al., 2020; Ellickson, 1994; Katchadourian, 2011) and between expectations of repayment or not (Ellickson, 1994; Gao et al., 2024). Integrating our three-component trust model with these more complex emotional constructs may prove useful, as it is likely that these constructs of inequity aversion, shame, guilt, blame, forgiveness, indebtedness and expectations interact differently with the different kinds of trust. Direct expectations of indebtedness likely impact deliberative systems, while constructs of inequity aversion and guilt likely impact more instinctual systems (Bellucci et al., 2017; Chang et al., 2011; Eisenberger, 2012; Gao et al., 2024; Sanfey et al., 2003). Both of these impact the fundamental question of reputation.

Several researchers have identified trust with the question of reputation specifically. The question of reputation is a particularly important one for trust in that it encodes much of the belief structure involved in the expected reduction of risk (Alexander, 2010; Axelrod, 1986; Boehm, 2012; Dunbar, 1996; Emler, 1994; Redish, 2022). Typically, these models look at the question of priors separately from that of experiential learning, and model the experiential learning as a reinforcement learning question, particularly in the context of the investment or trust game (Chang et al., 2010; Fouragnan et al., 2013). These models find that reputation priors are often encoded in prefrontal structures associated with our hypothesized deliberative system, but that experiential reputation is encoded in striatal structures associated with our hypothesized procedural system (Bellucci et al., 2020; Fouragnan et al., 2013). These priors are typically provided as explicit information about the “trustworthiness” of the partner. When the priors are instead provided as signals about the group-relationships of the partner, signals show up in other areas (anterior insula, amygdala) (Eisenberger, 2012; Engell et al., 2007; Singer, 2008; Todorov & Engell, 2008). Our three-system hypothesis suggests that explicit information would affect deliberative trust, in-group/out-group information would affect instinctual trust, and slowly experienced wins and losses would affect procedural trust.

Importantly, several psychiatric disorders such as psychopathy and borderline personality disorder can lead to difficulties in these interactions and thus to an inability to create and maintain trust (Driessen et al., 2021; Gong et al., 2019; Kazinka et al., 2024; King-Casas et al., 2008; Marsh, 2017; Wisner et al., 2021). Whether this is due to failures of instinctual systems and empathy (Marsh, 2017; Singer, 2008), an inability to recognize reciprocity (King-Casas et al., 2008), or an inability to express guilt at failures (Driessen et al., 2021) is still debated. Similarly, psychosis often includes paranoia, creating expectations of betrayal, even against evidence and under conditions in which the subject themself does not express betrayal (Kazinka, 2022; Kazinka et al., 2024). It would be very interesting to identify individual differences in instinctual failures (lack of empathy, difficulties in community construction), deliberative failures (changes in expectation, ability to represent or use those predictions), and procedural failures (inability to learn from rewards and consequences) may show up differently in different subjects. It is possible that looking at these psychiatric conditions as changes in information processing leading to changes in computational processing (Anticevic & Murray, 2017; Huys et al., 2016; Redish et al., 2008, 2022; Redish & Gordon, 2016; Series, 2020) may provide important insights into the mechanisms of trust.

Two other commonly used constructs integrate especially well with our framework: Edmondson's psychological safety (Edmondson, 1999; Edmondson & Lei, 2014) and Simons's behavioral integrity (Simons, 2008). Psychological safety arises in teams that perceive their environment as supportive of interpersonal risk-taking; for example, when nurses feel safe admitting mistakes. It is a non-dyadic phenomenon shaped by outcomes and experiences rather than traits of specific individuals, and its proposed strategies align with the learning mechanisms described in our model. Like our account, it also predicts that errors can serve as opportunities to build trust. Behavioral integrity, defined as the perceived consistency of words and actions, similarly fits with a computational perspective: consistency reduces the cognitive cost of deciding to cooperate, regardless of the specific values espoused. Integrating such organizational constructs with neuroscience-based decision models may provide a promising direction for future research.

## Limitations

In the interests of space, we have not addressed all of the neural functioning in the brain. Dopamine and other neuromodulators (such as norepinephrine, acetylcholine, serotonin, enkephalins, endorphins, dynorphin, endocannabinoids) are clearly important for neural function and decision-making in general, and social functioning in particular (Aston-Jones & Cohen, 2005; Dalley et al., 2004; Doya, 2002; Knutson & Greer, 2008; H. Steiner & Gerfen, 1998; Watson & Tellegen, 1985). Changes in these neuromodulators are very likely to impact trust behaviors. How exactly these neuromodulators interact with the various decision systems is currently controversial, and, thus, we believe better left for later discussion. Similarly, there is an extensive literature on the role of oxytocin in trust behaviors (Macdonald & Macdonald, 2010; Neto et al., 2020; Zak et al., 2007), the main effect of which seems to be to focus ingroup and outgroup dynamics, strengthening ingroup trust, but also outgroup distrust (De Dreu & Kret, 2016). If true, that would suggest a strong role for oxytocin in instinctual systems (Redish, 2022), but whether it has similar or other effects on deliberative systems (which include evaluative steps that may access affective components (Damasio, 1996; Shenhav, 2024)) or procedural systems (which do not) remains unknown.

As noted above, although the trichotomy of deliberative, instinctual, and procedural decision-making processes are components of nearly every computational analysis of decision making, they are not the only ones that have been proposed. There are a number of other potential computational decision-making systems that have been suggested, including systems based on linguistic logic (Newell & Simon, 1972), active inference (Parr et al., 2022), working memory (Collins, 2026), and simple action repetition (Miller et al., 2019). Similarly, some researchers have suggested additional systems that have mixed characteristics of the three primary systems, such as successor representations and chunking that treats environments procedurally, but assesses outcomes deliberatively (Dezfouli & Balleine, 2012; Geerts et al., 2019; Momennejad et al., 2017; Mugan, Amemiya, et al., 2024; Newell, 1992), or model-based Pavlovian (instinctual) learning that uses explicit expectations of outcomes with a limited repertoire of actions (Dayan & Berridge, 2014; Rescorla, 1988). In the interests of space, we have not tried to include their consequences for trust. Nevertheless, we note that the same computational analysis we have laid out here can be used to factor those in as well. Similar institutional consequences could be identified that could impact those systems. Perhaps testing institutional consequences could be used to demonstrate and validate those additional systems.

## Conclusion

Trust is fundamental to human societies, yet its dynamics remain difficult to predict. We argue that predictive power comes from grounding trust in the same neural systems and learning mechanisms that govern all forms of adaptation. What makes trust different from other expectations is not the information processing algorithms, but the vulnerabilities in the situations and the entanglement of those vulnerabilities with all the other demands of the environment. Our taxonomy identifies three complementary forms of trust — deliberative, instinctual, and procedural — each rooted in distinct decision-making systems and distinct information processing mechanisms. This new taxonomy suggests major modifications to existing notions of rational and emotional forms of trust. It also introduces procedural trust, which to our knowledge is not found in any existing taxonomy of trust. By correctly understanding the fundamentals of information representation, information processing, and learning mechanisms within the physical brain, we can design environments and institutions such that errors and violations of trust generate opportunities to build trust.

This work provides a new path forward to address institutional and systemic mechanisms to engender trust. It also provides predictions about how those systemic changes will affect trust within communities, and thus provides new ways to create and repair trust in organizations such as businesses, classrooms, and societies, as well as new ways to address the complex question of trust in social media and AI technologies (Bonnefon et al., 2024; Cushman, 2024; Glikson & Woolley, 2020; Redish, 2022; Reeves & Nass, 1996; Van Bavel et al., 2024). We suggest that aligning institutional and systemic mechanisms with the computational processes of decision-making (and, in particular, the learning processes therein) will increase their effectiveness to enable those institutions to work better together in cooperative communities based on trust.

## Contributions and conflicts of interest

All three authors co-wrote the paper. The authors declare that they have no conflicts of interest related to the work in this paper.

## Funding

This work was partially funded by NSF 2237593, the Jacobs Foundation, as well as resources from Cornell University and the University of Minnesota.